\documentclass[a4paper,11pt]{article}
\pdfoutput=1 

\usepackage{jheppub} 

\usepackage[T1]{fontenc} 

\usepackage{amsmath}
\usepackage{dsfont}
\usepackage{epsfig}
\usepackage{slashed}
\usepackage{bbold}
\usepackage{psfrag}
\usepackage{color}
\PassOptionsToPackage{caption=false}{subfig}
\usepackage{subfig}
\usepackage{multirow}
\usepackage{booktabs}
\usepackage{hyperref}
\usepackage{pstricks}
\usepackage{feynmp}
\usepackage[utf8]{inputenc}
\usepackage{cleveref}

\newcommand{\beq}{\begin{equation}}
\newcommand{\eeq}{\end{equation}}
\newcommand{\bea}{\begin{eqnarray}}
\newcommand{\eea}{\end{eqnarray}}

\newcommand{\FdisE}{\ifmmode {\lvert F_{dis}(E_r,E_{int}) \rvert}^2 \else ${\lvert F_{dis}(E_r,E_{int}) \rvert}^2$\fi}
\newcommand{\Fmol}{\ifmmode {\lvert F_{mol}(\mathbf{q,\tilde{\mathbf{q}}}) \rvert}^2 \else ${\lvert F_{mol}(\mathbf{q},\tilde{\mathbf{q}}) \rvert}^2$\fi}

\newcommand{\Mfi}{\ifmmode {\lvert \mathcal{M}_{2-2} \rvert}^2 \else ${\lvert \mathcal{M}_{2-2} \rvert}^2$\fi}

\renewcommand{\subsubsection}[1]{\addtocounter{subsubsection}{1}
\par\nobreak
\medskip
\nobreak
\noindent{\it \thesubsubsection.  #1 }
\par\nobreak\medskip\nobreak}

\def\lpar#1#2#3#4{\rlap{\raise#3\hbox{$\hskip#4#1\left\{\mbox{\phantom{\rule[0mm]{0mm}{#2}}}\right.$}}}
\def\rpar#1#2#3#4{\rlap{\raise#3\hbox{$\hskip#4\left\}#1\mbox{\phantom{\rule[0mm]{0mm}{#2}}}\right.$}}}

\begin{document}

\title{Accretion of Dissipative Dark Matter onto Active Galactic Nuclei}

\affiliation[a]{Raymond and Beverly Sackler School of Physics and Astronomy, Tel-Aviv University, Tel-Aviv 69978, Israel}
\affiliation[b]{School of Natural Sciences, Institute for Advanced Study,Einstein Drive, Princeton, NJ 08540, USA}
\affiliation[c]{Princeton Center for Theoretical Science, Princeton University, Princeton, NJ 08544, USA}
\affiliation[d]{Department of Physics, Loyola University Chicago, Chicago, IL 60660, USA}
\affiliation[e]{INFN - Sezione di Trieste, Via Bonomea 265, 34136, Trieste, Italy}
\affiliation[f]{SISSA International School for Advanced Studies, Via Bonomea 265, 34136 Trieste, Italy}

\author[a,b]{Nadav Joseph Outmezguine,}
\author[c]{Oren Slone,}
\author[d]{Walter Tangarife,}
\author[e,f]{Lorenzo Ubaldi,}
\author[a,b]{Tomer Volansky}

\emailAdd{nadav.out@gmail.com}

\emailAdd{oslone@princeton.edu}

\emailAdd{wtangarife@luc.edu}

\emailAdd{ubaldi.physics@gmail.com}

\emailAdd{tomerv@post.tau.ac.il}

\abstract{
We examine the possibility that accretion of Dissipative Dark Matter (DDM) onto Active Galactic Nuclei (AGN) contributes to the growth rate of Super Massive Black Holes (SMBHs). Such a scenario could alleviate tension associated with  anomalously large SMBHs measured at very early cosmic times, as well as observations that indicate that the growth of the most massive SMBHs  occurs  before $z\sim6$, with little growth at later times. These observations are not readily explained within standard AGN theory. We find a range in the parameter space of DDM models  where we both expect efficient accretion to occur and which is consistent with observations of a large sample of measured SMBHs. When DDM accretion is included, the predicted evolution of this sample seems to be more consistent with assumptions regarding maximal BH seed masses and maximal AGN luminosities. 
}

\maketitle
\flushbottom

\newpage

\section{Introduction} 
\label{sec:Intro}

Several well established observations indicate that Dark Matter (DM) behaves mostly as a cold and collisionless gas. 

A combination of theoretical prejudice, related to the hierarchy problem of the Standard Model of particle physics, together with the principle of Occam's razor, has traditionally led to consider DM as a single, weakly-interacting,  gravitating particle. In the past few decades, most theoretical and experimental efforts have centered around this kind of particle, largely overlooking other possibilities.  

The situation has changed in recent years.  With the lack of experimental evidence for DM or for a solution to the hierarchy problem, light DM residing in a low-scale ``dark sector'' has attracted significant attention~\cite{Boehm:2003hm, Boehm:2003ha, An:2014twa,An:2013yua, Feng:2008ya, Hooper:2008im, Fischler:2010nk, DAgnolo:2015ujb, Kusenko:2009up, Kaplan:2009ag, Essig:2010ye, Choi:2011yf, Falkowski:2011xh, Lin:2011gj, Boddy:2014yra, Rajagopal:1990yx, Covi:1999ty, Falkowski:2017uya}. The low, sub-GeV, mass scale arises both theoretically and observationally:  from the theoretical viewpoint, many production mechanisms that explain the observed relic abundance, require DM to have a low mass and possibly to interact strongly within the dark sector (see, e.g.,~\cite{Hochberg:2014dra,Hochberg:2014kqa,Knapen:2017xzo,DAgnolo:2018wcn,DAgnolo:2017dbv,Finkbeiner:2007kk,Pospelov:2007mp,Bernal:2015bla,DAgnolo:2015nbz,Kuflik:2015isi,Dror:2016rxc,Kopp:2016yji});  from the observational perspective, several discrepancies with N-body simulations suggest that DM may have rather strong self-interactions~\cite{Spergel:1999mh,Tulin:2013teo,Kaplinghat:2015aga,Boehm:2001hm}. Such interactions are likely to be mediated by light states, pointing again to the possibility of a strongly, self-interacting  light dark sector.

The presence of a complex, low mass dark sector not only predicts self-interactions, but often allows for multiple DM states, some of which may experience significant dissipative forces due to the emission of light particles (such as dark photons).   
While only a small fraction of the DM, $\lesssim 5\%$~\cite{Fan:2013yva}, can be dissipative, it has been shown that such a component may result in interesting phenomenological and measurable consequences. 
So far, dissipative DM has been studied mostly in relation to galactic structure~\cite{Fan:2013yva, Fan:2013tia,Foot:2013vna, McCullough:2013jma, Fan:2013bea, Foot:2014mia, Randall:2014kta, Heikinheimo:2015kra, Kramer:2016dqu, Buckley:2017ttd} and CMB signatures~\cite{Cyr-Racine:2013fsa, Buckley:2014hja}.
It is natural to ask whether one can constrain or discover dissipative DM using small-scale structures.

In this paper, we make progress in this direction by studying 
Active Galactic Nuclei (AGN).
AGN  are currently understood as Super Massive Black Holes (SMBHs) at the centers of galaxies, which are undergoing an active phase of accretion of matter~\cite{2013peag.book.....N}.
It is believed that baryonic matter 
forms an accretion disk that surrounds the BH. 
By losing angular momentum, this matter falls into the BH, feeding it, while at the same time releasing radiation to the surroundings. Although the accretion mechanism has been investigated for the past few decades (for an introduction and references, see for example~\cite{2002apa_book_Frank,Blaes:2007zm,2013peag.book.....N,2001ApJ...559..680H}), accretion disks are not well understood~\cite{1989ApJ...337..236C,1991A&A...248..389C,1979ApJ...227L..55S,2017MNRAS.467.3723K,Abramowicz:2011xu}. In particular, a first principle understanding of the origin of the viscosity, necessary for angular momentum loss and consistent with maintaining a steady state disk, is still lacking~\cite{Shakura:1976xk, 1981ARA&A..19..137P, Loeb:1991cb}. Instead, a phenomenological approach is often used, utilizing the so called $\alpha$-disk prescription~\cite{Shakura:1972te}, which encapsulates in a single parameter our ignorance about the microscopic properties of the viscosity. 

Independent of the detailed understanding of accretion disks, some interesting conundrums arise observationally. In particular, standard accretion models seem to be in tension with the fact that some BHs, in AGN observed at redshifts $z \sim 4$ to 7, are more massive than $10^9 \ M_\odot$~\cite{Trakhtenbrot:2010hj, Mortlock:2011va, Venemans:2012dt}.
Known mechanisms for BH formation at $z \sim 20 - 30$ allow a maximal seed mass of about $10^6 \ M_\odot$ ~\cite{Alexander:2011aa, Latif:2016qau} (see discussion below).
Despite the exponential rate of accretion predicted by the simple scenario where $\dot M_{\rm BH} \propto  M_{\rm BH}$, such accretion still fails to grow the largest observed SMBHs fast enough to match their measurements at $z \sim 4 - 7$. This discrepancy could conceivably be relaxed by taking into account merger events~\cite{Haiman:2004ve}, or assuming periods of super-Eddington growth~\cite{Madau:2014pta}. However, it remains
unclear whether these options provide a satisfactory solution to the observations. Therefore, it is of interest to explore alternative scenarios.

The presence of a strongly, self-interacting dark sector, provides the possibility of addressing the issue outlined above even if this sector is only a subdominant component of the total DM mass density\footnote{The requirement of a subdominant component allows this self-interacting DM to evade current bounds from other measurements.}. If the DM in such a sector has ultra-strong self interactions ($\sigma / m_{\rm DM} \sim 10^5 - 10^7 \ {\rm cm}^2 / {\rm g}$), it could seed much larger BH masses at early redshifts via gravothermal collapse~\cite{Pollack:2014rja}. Furthermore, if this DM component is also dissipative, and for more moderate interaction strengths ($\sigma / m_{\rm DM} \sim 1 \ {\rm cm}^2 / {\rm g}$), it can contribute to the growth rate of SMBHs, just as does baryonic matter, and as was mentioned recently in~\cite{DAmico:2017lqj}. We further explore the latter option in this study, under the assumption of a dark sector that is sufficiently similar to our visible sector such that disk formation and viscous accretion might be expected to behave similarly in both sectors. Under this assumption, we identify the additional necessary conditions that the dark sector must satisfy in order to contribute to SMBH accretion. We find regions in the dark sector parameter space which relieve some of the tension described above and which are consistent with SMBH mass observations. Furthermore, we find that for some regions within the consistent parameter space, accretion in the hidden sector is expected to occur mainly at high redshifts ($z\gtrsim 6$), a feature which is in agreement with recent observations~\cite{2016ApJ...819..123N} and which remains largely unexplained in the standard accretion scenario. 
This study lays the ground to a more detailed analysis in the future, and offers a new direction to probe and discover dissipative dark matter.

The paper is organized as follows.  In Sec.~\ref{sec:AGNs} we briefly review the basic physics of BH accretion and describe the current observational status.  In Sec.~\ref{sec:Model} we discuss a simple toy model for dissipative DM. In Sec.~\ref{sec:DMAccretion} we study the conditions for the formation of a dissipative DM accretion disk, and quantify the contribution from such a disk to the BH growth rate. We conclude with some brief remarks in Sec.~\ref{sec:Conclusions}.

\section{Supermassive Black Holes: Observations and Challenges} 
\label{sec:AGNs}

SMBHs are believed to be the result of accretion onto ``seed'' BHs, which are formed at high redshift~\cite{Volonteri:2010wz, Latif:2016qau}. There exist three main candidate mechanisms for seed formation: 
(i) remnants of population III stars \cite{Heger:2001cd} resulting in BH seed masses, $M_{\rm seed}$, in the range of 10 -- 100 $M_\odot$, at $z\simeq 20 - 50$;
(ii) direct collapse of primordial gas clouds \cite{Begelman:2006db} resulting in masses as large as $M_{\rm seed} \sim 10^6  \ M_\odot$\footnote{Although the upper limit of $M_{\rm seed} \sim 10^6  \ M_\odot$ is achievable via this mechanism, the same study~\cite{Begelman:2006db} predicts that not all such collapses terminate with a SMBH of maximal mass. Thus, it should be expected that many SMBH seeds are far less massive.}, at $z\simeq 5 - 10$; (iii) merging of dense stellar clusters \cite{Bernadetta:2008bc} resulting in masses up to $M_{\rm seed} \sim 10^3 \ M_\odot$, at $z\simeq 10 - 15$. As material falls into the BHs, these seeds grow and, by redshift $z < 7$, some of them are observed to have acquired masses larger than $10^9 \ M_\odot$~\cite{Trakhtenbrot:2010hj, Page:2013zoa, Fan2006}. Currently, there is no completely satisfactory explanation of the accretion process. 

There are a number of mechanisms for accretion of matter onto a SMBH. In its simplest form, gas particles that have a velocity smaller than the escape velocity from the BH, fall towards the BH in the radial direction. This is known as Bondi accretion~\cite{Bondi:1952ni}, but is inefficient when the infall process must also overcome an angular momentum barrier. It is therefore important to have a mechanism for the removal of angular momentum in order to create a substantial BH growth rate.
Even though a detailed understanding of such a mechanism is still lacking, it is widely believed that accretion disks form around BHs in AGN and fuel their growth. A full understanding of the viscosity is still lacking and it is common to introduce a parameter, $\alpha$, that encodes the unknowns in the microphysics of the gas and relates the viscosity to macroscopic parameters: the height of the disk and the speed of sound. The resulting family of models are referred to as $\alpha$-disk models~\cite{Shakura:1972te}.

\subsection{Basic AGN Concepts} 
\label{sub:basic_agn_concepts}
A simple picture consists of a thin accretion disk, with a sub-parsec radius. The disk contains ionized gas which, due to viscosity, loses angular momentum and falls towards the BH. As matter falls into the BH, part of its gravitational potential energy is converted into radiation, which is observed as the AGN luminosity, 
\beq\label{eq:L_Mdot}
L = - \eta \dot{M}_{\rm disk} \, .
\eeq
Here $\eta$ is the radiative efficiency, which ranges from 0.057 to 0.42~\cite{Shapiro:2004ud} and depends solely on the BH geometry, $M_{\rm disk}$ is the disk mass, and the overdot denotes a time derivative. 
Conventionally, $L$ has an upper limit, the Eddington luminosity, corresponding to the configuration in which the outward radiation pressure equals the inward gravitational pull\footnote{This expression is obtained under the assumption of spherical accretion, which is not the case for accretion disks. However, observed AGN luminosities rarely exceed the Eddington limit, and even then, by no more than an order one factor.  This result is therefore commonly used in the literature and provides a conservative limiting rate at which accretion can occur even in the case of disk geometries.},
\begin{equation} \label{eq:L_edd}
L_{\rm Edd}= 4 \pi G_N\frac {M_{\rm BH}m_{p}}{\sigma_{ T}} \, ,
\end{equation}  
where $G_N$ is Newton's constant, $M_{\rm BH}$ the BH mass, $m_{p}$ the proton mass and $\sigma_{ T}$ the Thomson scattering cross section. The BH accretion rate can be written as
\beq
\dot M_{\rm BH} = -(1- \eta) \dot M_{\rm disk} = \frac{1-\eta}{\eta} \frac{L}{L_{\rm Edd}} \frac{M_{\rm BH}}{\tau_{\rm Sal}} \, ,
\label{eq:Mdot_L}
\eeq
where
\begin{equation}
	\tau_{\rm Sal} \equiv \frac{\sigma_T}{4\pi G_N m_ {p}} \simeq 4.5 \times 10^8 \ {\rm yr}
\end{equation}
is the Salpeter time~\cite{Salpeter:1964kb}. The accretion is not necessarily continuous. In general, there are active and inactive phases for a galactic nucleus, called duty-cycles. In this study, we are mostly interested in the long time behavior, in which case these phases can be taken into account by using a time-averaged growth rate,
\begin{equation}
\langle\dot M_{\rm BH}\rangle =  \frac{1-\eta}{\eta} \frac{L}{L_{\rm Edd}} \frac{M_{\rm BH}}{\tau_{\rm Sal}} D\,,
\label{eq:Mdot_L_avg}
\end{equation}
where $0<D<1$ is the (time averaged)  duty cycle. Its value can be inferred statistically, at various redshifts, by measuring the BH mass distribution of active galaxies at different cosmic epochs. 
Under the assumptions of $\eta=0.1$ and either constant $L$ or constant $L/L_{\rm Edd}$, measurements point
 towards values of $D$ on the order of $\mathcal{O}(0.1)$ at $z\lesssim6$~\cite{Trakhtenbrot:2010hj}. 

Studies show that the value of $L / L_{\rm Edd}$ could change as a function of redshift and BH mass. Specifically, it is predicted that there could be an anti-correlation between $M_{\rm BH}$ and $L / L_{\rm Edd}$~\cite{0004-637X-648-1-128,Shankar:2007zg,2004MNRAS.353.1035M,2004MNRAS.352.1390M,2007ApJ...654..754N}; however, a common assumption in the literature is that of a constant $L / L_{\rm Edd}$~\cite{KING2008253,Trakhtenbrot:2010hj}. Under this assumption, the luminosity (and therefore the accretion) scales with SMBH mass, resulting in an exponential growth rate of the mass. 
Consequently, the time for a BH to grow to a mass $M_{\rm BH}$ from the initial $M_{\rm seed}$ follows from Eq.~\eqref{eq:Mdot_L_avg},
\begin{equation}\label{Eq:baryonic_growth}
	t_{\rm grow} = \frac{\tau_{\rm Sal}}{D}\frac{\eta}{ 1-\eta}\left(\frac{L}{ L_{\rm Edd}}\right)^{-1} \ln \left( \frac{M_{\rm BH}}{M_{\rm seed}} \right) \, .
\end{equation}
From Eq.~\eqref{Eq:baryonic_growth}, one can derive an upper bound on the growth rate by taking $D=1$.


\subsection{AGN Observations} 
\label{sub:agn_observations}

\begin{figure}
\begin{center}
\includegraphics[width=.47\textwidth]{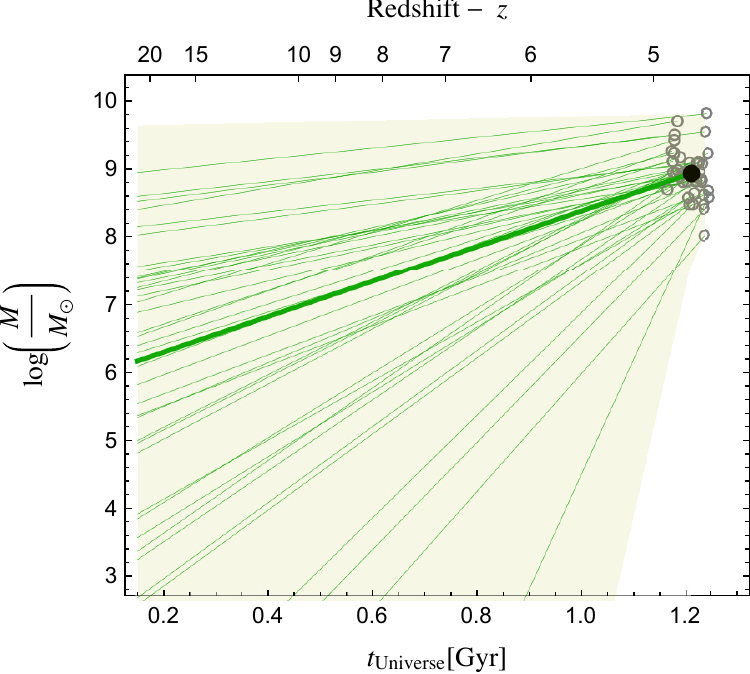}\;\;\;\;\;\;\includegraphics[width=.46\textwidth]{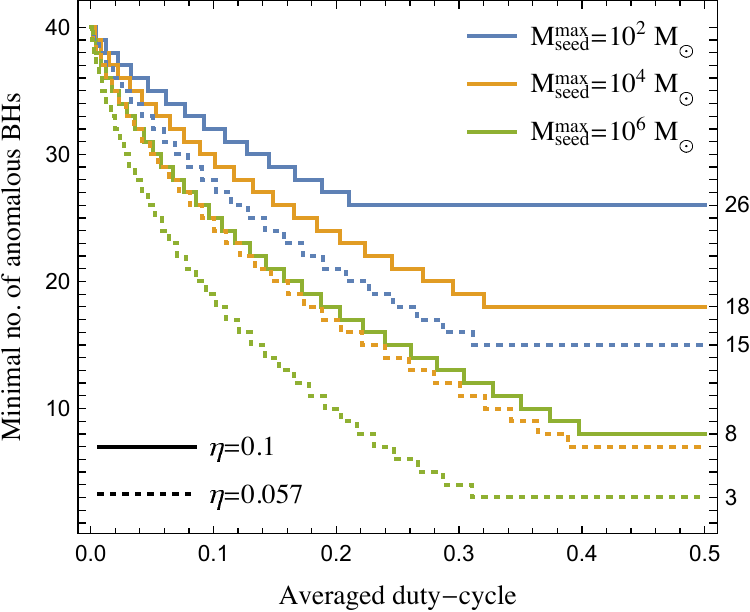}
\caption{\label{FIG:Netzer} 
{\bf Left panel}: Time evolution of BH masses according to Eq.~\eqref{Eq:baryonic_growth}, 
under the assumption of $\eta=0.1$ and duty cycle $D=0.5$, for the sample of 40 AGN analyzed in Ref.~\cite{Trakhtenbrot:2010hj}. The \textit{gray circles} represent the measured BH masses. 
The \textit{black dot} is the sample's mean mass. 
The {\it thin green curves} show the evolution of each SMBH in the sample. Their slopes are dictated by the ratios $L/L_{\rm Edd}$ measured for each BH, which are assumed to be constant throughout the evolution. The \textit{thick green curve} corresponds to the mean value of the sample, $L/L_{\rm Edd} \approx 0.6$. Note that, in this simple picture, some BHs require seed masses much larger than $10^6 \ M_\odot$ at $z = 20$. The \textit{yellow shaded region} represents the region enclosing all possible BH-growth histories if the duty cycle is allowed to vary between $0.1$ and $1$. {\bf Right panel}: Minimal number of BHs in the sample that would require ``anomalously'' fast growth, i.e. at a rate faster than that predicted by Eq.~\eqref{Eq:baryonic_growth}, as a function of the averaged duty cycle. Different colors correspond to different seed masses at $z = 20$. \textit{Solid curves} correspond to the reference value of accretion efficiency $\eta=0.1$, while \textit{dotted curves} correspond to the minimal possible value of $\eta=0.057$. 
}
\end{center}
\end{figure}

There are a number of methods to measure SMBH masses in distant galaxies.
One of these methods, known as reverberation mapping, is based on spectroscopic data rather than on total AGN luminosity and is considered to have relatively lower uncertainties. A large number of such luminosity-independent BH mass measurements were performed in Refs.~\cite{Netzer:2004pi, Kurk:2007qk, Netzer:2007is, Willott:2010yu, Trakhtenbrot:2010hj}. The study of Ref.~\cite{Trakhtenbrot:2010hj} provides a sample of 40 AGN, measured at redshift $z\sim4.8$. In this sample, the mean BH mass is $\sim 8\times10^8M_\odot$ and the mean value for $L/L_{\rm{\rm Edd}}$ is $\sim 0.6$. 
Under the assumption of constant $L / L_{\rm Edd}$ and setting $\eta=0.1$, the authors found that many of the measured SMBHs require extremely large seed masses. Specifically, if one considers the most optimistic (albeit unlikely) scenario with duty cycles of unity, $D=1$, and $\eta= 0.057$, then 7 out of the 40 AGN which were studied require a seed mass larger than $M_{\rm seed} > 10^4 \ M_\odot$ at $z = 20$.  Such a fraction of very massive seeds is above the predictions of the mechanisms mentioned earlier. In a more realistic scenario, in which each AGN has a different duty cycle, one can calculate the minimal number of anomalous SMBHs as a function of the average duty cycle. Results for such a calculation, for the same sample of SMBHs, are shown on the right panel of Fig.~\ref{FIG:Netzer}. As is evident in the figure, increasing the maximal seed mass or the average duty cycle (or a combination of these), reduces the minimal number of anomalous SMBHs. Furthermore, for a given maximal seed mass, there is a certain duty cycle above which the number of anomalous SMBHs remains constant. This happens since some of the BHs are anomalously heavy independently of the value of $D$. The left panel of the same figure shows the measured data, together with the calculated evolution of the SMBHs' masses over time, assuming constant $L / L_{\rm Edd}$, $\eta=0.1$ and $D=0.5$.  

\

This picture of accretion with efficiency $\eta$ and a constant duty cycle is rough and possibly optimistic as it assumes constant 
average growth from seed formation until late redshifts. This is probably not true. The accretion process is likely to occur in bursts with varying duty cycles. Specifically, some recent studies suggest that the most massive SMBHs experience most of their growth at very early epochs ($z\lesssim 6$)~\cite{2016ApJ...819..123N,2014ApJ...791...34N}. Thus, the situation of anomalously large SMBHs, as described above, could be even more severe because one must additionally explain the existence of both very heavy and very early-formed SMBHs.  As mentioned before, some caveats, such as possible periods of super-Eddington accretion or additional growth through BH-BH mergers, can ameliorate the tension present in the naive picture  discussed above. Nevertheless, it is worthwhile to explore other avenues in order to address the puzzle.
In this work, we study how dissipative dark matter, with weaker self interactions compared to those needed in the mechanism of~\cite{Pollack:2014rja}, could form a dark accretion disk and contribute to fueling the growth of SMBHs.

\section{Dissipative Dark Matter}
\label{sec:Model}

In this section, we present a concrete and simple model of the hidden sector. This model will be used for reference in the rest of this study, however, it should be emphasized that the discussion in the sections to follow is generic and applies also to other (possibly more complex) models. The only requirements of such models is that they allow for efficient dissipation and efficient cooling for a large enough fraction of mass within the hidden sector.

The example model we consider here involves a hidden sector which contains two different DM components. The dominant of these components is non-dissipative, cold DM (CDM). 
The second component is composed of two fermions $p^\prime$ and $e^\prime$, with a mass hierarchy $m_{p^\prime} \gg m_{e^\prime}$. These fermions have opposite charges under some unbroken, hidden $U(1)$ gauge group, with fine structure constant $\alpha^\prime$. This is a simplified model mimicking QED with protons and electrons, which allows for dissipation in the dark sector. We refer to it as dissipative dark matter\footnote{The authors of Ref.~\cite{Fan:2013yva} refer to the same toy model as double-disk dark matter (DDDM).} (DDM). We introduce no mass for the Abelian vector boson since we are interested in a long range interaction. For simplicity, we also neglect the possibility of kinetic mixing between the hidden vector and the SM photon. Numerous studies of this sort of hidden sector have been presented in the literature \cite{Goldberg:1986nk,Mohapatra:2001sx,Kaplan:2009de,Behbahani:2010xa,CyrRacine:2012fz,Cline:2013pca,McCullough:2013jma,Fan:2013bea,Banks:2014rsa,Fischler:2014jda,Foot:2014uba,2014PhRvL.112p1301R,Randall:2014kta,Reece:2015lch,Kramer:2016dqu,Kramer:2016dew,Agrawal:2017rvu,Rosenberg:2017qia,DAmico:2017lqj,Buckley:2017ttd}. 
As will be shown below, additional (possibly short-range) interactions between the hidden proton and electron may be necessary in order for the cooling process to be efficient.  As a simple illustration of this, we consider 
the presence of an additional massive mediator, $\phi$, with mass $m_\phi$ and coupling $\alpha_\phi$.

The relic abundance of the charged heavy particles, $p^\prime$ and  $\bar p^\prime$, receives a minimal contribution from the freeze-out~\cite{Feng:2008mu} of annihilations into two hidden  photons and into dark electron-positron pairs. We require that such an abundance is, at most, a $5\%$ fraction of the total DM relic density. This bound is well within the limit derived from merging of galactic clusters~\cite{Markevitch:2003at}, and simultaniously complies with the Oort limit, which constrains the local dark matter density; the latter has been studied in Refs.~\cite{Fan:2013yva, Schutz:2017tfp} for the specific case of DDM under consideration here. For a given gauge coupling and under the assumption of thermal production, this constraint implies an upper limit on $m_{p'}$. We stress that this limit is model dependent. For example, the addition of extra light degrees of freedom into which the symmetric component can annihilate would reduce the freeze-out abundance of $p^\prime$ and  $\bar p^\prime$. Regarding $e^\prime$ and $\bar e^\prime$, being much lighter, their freeze-out relic abundance is negligible compared to that of the dark protons. An additional production mechanism is required in order to set an asymmetry $n_{e^\prime}\,- \,n_{\bar e^\prime} = n_{p^\prime}\,- \,n_{\bar p^\prime} >0$. A large variety of such mechanisms can be found in the literature \cite{Nussinov:1985xr,Kaplan:1991ah,Kaplan:2009ag,Kribs:2009fy,Shelton:2010ta,Davoudiasl:2010am,Falkowski:2011xh,MarchRussell:2011fi,Cui:2011qe,Fischler:2014jda}.

A final assumption that we make is that the hidden and the visible sectors 
decouple sufficiently early in order to evade cosmological bounds on the number of relativistic degrees of freedom (see~\cite{Fan:2013yva} for details).

\section{Black Hole Accretion with Dissipative Dark Matter} 
\label{sec:DMAccretion}

Under the assumption that a dissipative hidden sector exists, it might be possible to enhance the growth rate of BHs by accretion of matter from this sector. 
For this to happen, a number of requirements must be fulfilled:

\begin{enumerate}

\item \textbf{Efficient accretion.} Assuming that enough DDM reaches the direct vicinity of the BH and forms a dark accretion disk, one needs an efficient viscosity mechanism that allows DDM to lose angular momentum and accrete onto the BH. As explained in Sec.~\ref{sec:AGNs}, such a mechanism is known to exist in the visible sector, even though it is not well understood. It is thus a reasonable assumption that a hidden dissipative sector, which is similar enough to the visible sector, has an analogous mechanism. When accretion does occur, it must be fast enough to allow the BH to grow efficiently, implying a value of hidden $\tau_{\rm Sal}$ small enough  to explain observations.

\item \textbf{Formation of a bound substructure.} Part of the DDM, originally distributed in the entire halo, must collapse and reach the direct vicinity of the BH. This happens as a consequence of cooling processes analogous to those in the visible gas. Whatever model the dark sector consists of, it is essential that these cooling processes transfer enough dark mass towards the SMBH.   
In the example discussed in the previous section, the hidden electrons, $e^\prime$, are cooled more rapidly than the hidden protons, $p^\prime$.  To ensure an efficient overall cooling, it is therefore essential that the two species remain in thermal equilibrium during the process. 
The total mass that is eventually bound in the dark accretion disk depends on the initial DDM density profile and on the radius of influence around the BH, i.e. the maximal radius at which a cold DDM particle cannot escape the BH's gravitational potential.  
We denote by $\tau_{\rm cool}$ the timescale for the DDM to reach the region within the BH's radius of influence. For any redshift, this cooling time must be shorter than the age of the universe at that redshift, $\tau_{\rm cool} \lesssim \tau_{\rm univ}$, but larger than the equilibration time, $\tau_{\rm eq} \lesssim \tau_{\rm cool}$, in order for accretion to ignite.

\item \textbf{Large duty cycles.} 
Once the DDM  substructure forms, it can accrete onto the BH. We denote by $\tau_{\rm acc}$ the timescale for the accretion of the mass within the bound substructure. This accretion timescale must be shorter than the age of the universe for any redshift, $\tau_{\rm acc} \lesssim \tau_{\rm univ}$. At the same time, the cooling timescale must be of order, or shorter than the accretion timescale, $\tau_{\rm cool} \lesssim \tau_{\rm acc}$, in order to allow for the formation of a long lasting dark disk (otherwise any formed disk will be accreted before additional mass reaches its vicinity). The dark duty cycle, $D^h$, should depend on the ratio $\tau_{\rm acc} / \tau_{\rm cool}$, and is expected to be unity when the rate of cooling is faster than the accretion rate, i.e. when $\tau_{\rm cool} < \tau_{\rm acc}$.

\end{enumerate}

We begin with a calculation of the BH growth rate requiring only point (1) and assuming that points (2) and (3) are automatically satisfied. The calculation is a simple generalization of that presented in 
Sec.~\ref{sec:AGNs} and is meant to provide an intuitive picture. We then consider the additional constraints arising from requirements (2) and (3) and study the resulting available parameter space in the  simple DDM model we consider. In what follows, we discuss these requirements in more detail.

\subsection{Accretion from an Existing Dark Accretion Disk}
\label{sec:naive}

Accretion disks are the most efficient way to accrete matter with net angular momentum onto a compact object such as a BH. In analogy with the visible sector, DDM, whose properties mimic those of baryons, can be expected to have similar viscous properties and to enhance the accretion rate. 
Following the simple prescription of Sec.~\ref{sec:AGNs}, the luminosity, $L$ from Eq.~(\ref{eq:L_Mdot}), now receives two contributions,
\beq
L \rightarrow L^{v}+L^{h},
\eeq
where the superscripts $v$ and $h$ correspond to the visible and hidden sectors respectively. Since the BH efficiency, $\eta$, depends solely on the BH spin~\cite{Shapiro:2004ud}, Eqs.~(\ref{eq:L_Mdot}) and (\ref{eq:Mdot_L_avg}) still hold. Extending the definition of the Eddington luminosity to the hidden sector, 
one can write the time averaged BH growth rate as
\begin{equation}
	\langle\dot{M}_{\rm BH}\rangle = \frac{1-\eta}{\eta} \left(D^v\frac{L^{v}}{L_{\rm Edd}^v} +\zeta D^h \frac{L^{h}}{L_{\rm Edd}^h} \right) \frac{M_{BH}}{\tau_{\rm Sal}},
\label{eq:MdotBH_vh}
\end{equation}
where we have introduced the duty cycles $D^{v}$ and $D^{h}$ in both sectors. $\zeta$ is the ratio of the Eddington luminosities in the respective sectors,
\begin{equation} \label{xidef}
\zeta \equiv \frac{L_{\rm Edd}^h}{L_{\rm Edd}^v} =\frac{\sigma_T/m_p}{\sigma^\prime_T/m_{p^\prime}},
\end{equation}
with $\sigma^\prime_T$ the hidden-sector equivalent of the Thomson cross section, $\sigma^\prime_T=(8 \pi/3)(\alpha^{\prime 2}/m_{e^\prime}^2)$.

Extending the standard assumption of exponential growth~\cite{Trakhtenbrot:2010hj} to the hidden sector, we take both $L^v/L_{\rm Edd}^v$ and $L^h/L_{\rm Edd}^h$ to be constant. Taking $M_{\rm BH}(t=0) = M_{\rm seed}$ and neglecting the effects of substructure, i.e. assuming the duty cycle is constant (we will relax this assumption below), Eq.~(\ref{eq:MdotBH_vh}) can be integrated to obtain
\begin{equation}
\log\left(\frac {M_{BH}}{M_{\rm seed}}\right)=\frac{(1-\eta)}{\eta} \left(D^v\frac{L^{v}}{L_{\rm Edd}^v} + D^h \zeta \frac{L^{h}}{L_{\rm Edd}^h} \right) \frac{t}{\tau_{\rm Sal}}.
\label{eq:Hidden_Accretion}
\end{equation}
This predicts the BH mass at time $t$, given the contribution of the DDM which is quantified by three parameters, $\zeta,\,D^h$ and $L^h/L_{\rm Edd}^h$. 

In accordance with Ref.~\cite{Trakhtenbrot:2010hj}, we set $\eta=0.1$, take values of the visible duty cycle between $0.1-1$, and begin by considering $M_\mathrm{seed}\leq 10^4M_\odot$ at redshift $z_\mathrm{seed}=20$. We further assume that the hidden accretion rate does not exceed the Eddington limit. Thus, the quantity $D^h L^h/L_{\rm Edd}^h$ falls in the range $0-1$. For each of the 40 measured AGN presented in Ref.~\cite{Trakhtenbrot:2010hj}, the ranges of the various parameters above yield a range of consistent $\zeta$ via Eq.~\eqref{eq:Hidden_Accretion}. The intersection of all these ranges is translated into a lower bound $\zeta\gtrsim 0.45$. Given the value $\sigma_T/m_p\simeq 0.4~ {\rm cm^2 / g}$ in the visible sector, the bound on $\zeta$ implies
\begin{equation}
	\frac{\sigma^\prime_T}{m_{p^\prime}}\lesssim0.9~\frac{\rm{cm^2}}{\rm g}
\label{eq:sigmaT_bound}
\end{equation}
in the hidden sector.
Allowing for yet larger seed BH masses, $M_{\rm seed} \leq 10^6 M_\odot$, the upper bound on $\sigma^\prime_T / m_{p^\prime}$ grows roughly by a factor of two.

It should be noted that Eq.~(\ref{eq:sigmaT_bound}) refers to the Thomson cross section for the scattering of dark photons off of dark electrons, not to be confused with the self-interaction cross section for DM-DM processes, $\sigma$, 
often studied in the literature. 
The latter, for $p^\prime - p^\prime$ scattering mediated by a dark photon, is $\sigma \sim (m_{e^\prime}/m_{p^\prime})^2 \sigma^\prime_T \ll \sigma^\prime_T$ and, combined with the fact that the $p^\prime$s constitute only a 5\% fraction of the total dark matter, has very little effect on the galactic structure. Thus, the resulting DM-DM scattering is of no concern for this study.

\subsection{Additional Conditions for the Formation of a Dark Accretion Disk}\label{sec:conditions}

Next, we address points (2) and (3) presented at the beginning of the section, namely the formation of a bound substructure, equilibration of DDM and the requirement of large duty cycles. We begin by considering an initial DDM density profile in hydrostatic equilibrium. This will be the basis for a conservative estimate of the DDM mass that can collapse down to form an accretion disk after cooling. We then estimate the timescales corresponding to cooling, equilibration and accretion, and assess the conditions that allow large duty cycles.

\subsubsection{The Initial DDM Density Profile}
\label{sub:galactic_toy_model}

In this section, we describe the configuration of the initial DDM galactic profile within the gravitational potential of the dominant CDM component. This provides an estimate of the amount of DDM mass in the vicinity of the central SMBH, which will be used below to determine the BH growth rate originating from DDM accretion. The DDM is expected to cool and collapse onto a disk from this initial spherical configuration. However, in this study we do not attempt to analyze the details of this collapse process, but instead only provide conservative estimates. 

We denote with $M_{\rm DM}^{\rm gal},\,M_{\rm DDM}^{\rm gal},\,M_{\rm CDM}^{\rm gal}$ the total DM, DDM and CDM mass in the galactic halo, respectively. Following the analyses of Refs.~\cite{Fan:2013yva,Fan:2013tia}, we assume that only a fraction,
\begin{equation}
\epsilon\equiv M_{\rm DDM}^{\rm gal}/M_{\rm DM}^{\rm gal}\sim 0.05,
\end{equation}
of the DM mass is dissipative. As explained in Ref. \cite{Fan:2013yva}, such a small fraction of DDM evades the constraints from halo shapes 
and cluster interactions\footnote{A stronger constraint on this fraction (of approximately 2\%) exists from dynamical measurements of stars in our local neighborhood~\cite{Schutz:2017tfp}. However, this constraint applies directly to the Milky Way only. Furthermore, it assumes a galactic dark disk which is coplanar with the stellar disk. Thus, this constraint may not hold if any non-coplanar DDM disk breaks up via dynamical friction.}. 
The remaining 95\% is in the form of CDM.
We assume that, before the DDM component cools down to form a disk, the configuration is spherically symmetric, with a CDM halo described by a Navarro Frenk White (NFW) profile~\cite{Navarro:1996gj},
and an isothermal spherical profile for DDM in hydrostatic equilibrium within the CDM gravitational potential. We neglect baryonic matter.

The NFW profile is parametrized as,
\begin{align}
& \rho_{\rm CDM} (r)  = \frac{\rho_{\rm CDM}^0}{\frac{r}{R_s}\left(1+\frac{r}{R_s}\right)^2}\; ,  \label{eq:NFW}\\
& \rho_{\rm CDM}^0 \,=\, \left(\frac{1- \epsilon}{3}\right)\left(\frac{C^3}{\log(1+C)-\frac{C}{1+C}}\right)\bar{\rho}\;, \qquad 
\bar{\rho}\equiv \frac{3M_{\rm DM}^{\rm gal}}{4\pi R_{\rm vir}^3} \,  \label{eq:NFWnorm} .
\end{align}
Here, $R_s$ is the scale radius, $C \equiv R_{\rm vir}/R_s$ the concentration parameter, $M_{\rm DM}^{\rm gal} \simeq M_{\rm vir}$, and we take the virial mass and radius to be $M_{\rm vir}=10^{12}M_\odot$ and $R_{\rm vir}=110 ~\rm kpc$ as reference values. Observations~\cite{Mandelbaum:2008iz} and simulations~\cite{Ludlow:2013vxa} suggest that for most galaxies, $C$ should range between $1-100$, where at low redshifts the values are larger, and at redshifts of $z > 5$, values decrease to $C\lesssim4$ ~\cite{Bullock:1999he}. 

The CDM plus DDM system is governed by the hydrostatic and Poisson equations. In App.~\ref{sec:ddm_profile_derivation}, we assume that the isothermal sphere is virialized and solve these equations self-consistently. This procedure allows one to derive the shapes of both the isothermal profile and the virial temperature as functions of the NFW concentration parameter. The resulting DDM profile is given by
\begin{align}
\rho_{\rm DDM} (r)  & =\rho_{\rm DDM}^0e^{-b}\left(1+\frac{r}{R_s}\right)^{b R_s/r} \, , \\
b &  = \frac{3}{C^{2}}\frac{\rho_{\rm CDM}^0}{\bar \rho} \frac{\bar T}{T_{\rm vir}}  \, , \label{eq:b_def}
\end{align}
where $\rho_{\rm DDM}^0 \equiv \rho_{\rm DDM}(r = 0)$ and $\bar T \equiv G M^{\rm gal}_{\rm DM} m_{p'}/2 R_{\rm vir}$. 

We mention here two results that are derived in App.~\ref{sec:ddm_profile_derivation} and are important for the reminder of this paper: (i) For a wide range of the concentration parameter, the calculated virial temperature is consistent with 
\begin{equation}\label{eq:T_vir}
	T_{\rm vir}\simeq \bar{T}=\frac{G M^{\rm gal}_{\rm DM} m_{p'}}{2 R_{\rm vir}}\,,
\end{equation}
up to an order one factor.
(ii) For $r\lesssim R_s$, the DDM profile is constant and equal to its central value $\rho_{\rm DDM}^0$. Hereafter, we adopt a concentration parameter $C=4$ for which
\begin{equation}\label{eq:rho_0}
	\rho_{\rm DDM}^0\simeq 4 \bar{\rho}=\frac{3M_{\rm DM}^{\rm gal}}{\pi R_{\rm vir}^3}.
\end{equation}

\subsubsection{Cooling and Substructure Formation}
\label{sub:cooling}

As the DDM gas cools, a substructure bound to the central SMBH is expected to form.
We define the volume of this bound DDM substructure as the region in which the thermal velocity, $v_{\rm th}$, of these particles is smaller than the escape velocity, $v_{\rm esc}$, of the BH+DM system. As this structure is formed, the gas cools down, and more regions of space around the SMBH become bound. Thus, the important parameter for cooling processes is the final temperature, $T_f$, discussed below. Solving $v_{\rm esc} = v_{\rm th}$, one finds the radius of the bound substructure around the SMBH to be
\begin{equation}\label{eq:r_sub}
	R^{\rm sub}_{\rm DDM}\simeq \frac{T_{\rm vir}}{T_f}  \frac{M_{\rm BH}}{ M^{\rm gal}_{\rm DM}}R_{\rm vir}.
\end{equation}
Since this radius is much smaller than $R_{\rm vir}$, the total DDM substructure mass can be estimated by taking a constant density within $R^{\rm sub}_{\rm DDM}$, which is equal to the central value $\rho_{\rm DDM}^0$,
\begin{equation}
	M_{\rm DDM}^{\rm sub}\simeq \frac{4 \pi}{3}\rho_{\rm DDM}^0\left(R^{\rm sub}_{\rm DDM}\right)^3\simeq  \frac{\rho_{\rm DDM}^0}{\bar{\rho}}\left(\frac{M_{\rm BH}}{M_{\rm vir}}\right)^2\left(\frac{T_{\rm vir}}{T_f}\right)^3M_{\rm BH}.
 	\label{eq:MDDM_MBH}
\end{equation}

Similarly to baryonic matter, the virialized light DDM particles, $e^\prime$, can cool down via inverse Compton scattering with the CMB (in this case the relic of dark radiation) and via bremsstrahlung radiation. Assuming that these light particles are as abundant as the heavier particles, i.e. that the number densities are equal $n_{e^\prime} = n_{p^\prime} \equiv n$, the cooling timescale due to these processes is given by (see App.~\ref{sec:temperature_evolution} for details)
\begin{align}
	&\tau_{\rm cool}\simeq \frac{2}{\Gamma_{\rm Comp}+u_0^{-1}\Gamma_{\rm brem}},\\
	&\Gamma_{\rm Comp}\simeq \frac{8 \pi}{45}\frac{\sigma^\prime_T}{m_{e'}}T_0'^4(1+z)^4\;\;,\;\;\Gamma_{\rm brem} \simeq\alpha'n\sigma^\prime_T\,,
\end{align}
where $u_0^2=T_{\rm vir}/m_{e'}$, $T^\prime_0$ is the dark CMB temperature today (which must be smaller than  roughly half of the CMB temperature of visible photons to evade cosmological constraints~\cite{Fan:2013yva}), $z$ is the redshift, and $\sigma^\prime_T$ is the dark Thomson cross section, defined below Eq.~\eqref{xidef}. Noting that $t_{\rm Comp}$ has a strong dependence on $z$, the prediction is that for much of the evolution of the SMBH, Compton cooling is the dominant process. These cooling processes remain efficient only for ionized particles; so the above holds as long as the DDM temperature is larger than the binding energy $E_B \approx \alpha^{\prime 2} m_{e^\prime} / 2$. However, the cooling processes are  not necessarily fast enough to allow for the gas to cool down to the binding energy within the time elapsed between $z\sim 20$ to $z\sim 4.8$ (formation to measurement, roughly a Gyr). We therefore estimate the final temperature of the gas as
\begin{equation}
	T_f\simeq\max\left[E_{B},T_{\rm vir}\exp\left(-\frac{t(z=4.8)-t(z=20)}{\tau_{\rm cool}}\right)\right].
\end{equation}

The processes described above provide efficient cooling for the light particles, $e^\prime$. If the heavy particles, $p^\prime$,
are  also to collapse down to smaller radii, they must follow the lighter companions. This happens if the 
$p^\prime$ and $e^\prime$ remain in thermal equilibrium as the latter cool down. The equilibrium can be maintained via Rutherford scattering of $e^\prime$ on $p^\prime$. The timescale for this equilibration can be studied by solving the Boltzmann equations presented in App.~\ref{sec:temperature_evolution}, from which one can estimate
\begin{equation}\label{eq:eq_timescale}
	\tau_{\rm eq} \simeq \frac{\Gamma_{\rm Comp}u_0^3+\Gamma_{\rm brem}u_0^2}{2\Gamma_{\rm Rut}}\, \tau_{\rm cool}\, ,
\end{equation}
with
\begin{equation} \label{GammaRut}
	\Gamma_{\rm Rut}\simeq\frac{m_{e'}n\, \sigma_{T'}}{\sqrt{8 \pi}m_{p'}}\log\left(\frac{32\, m_{e'}^2T_{\rm vir}^2u_0^4}{3 \pi\, n^2 \sigma_{T'}}\right).
\end{equation}
Our expression for $\tau_{\rm eq}$ agrees with Refs~\cite{1941ApJ....93..369S,SpitzerJr.:1942zz,Fan:2013yva}, and we have verified the validity of this approximation against the numerical solution of the Boltzmann equations.   
The requirement $\tau_{\rm eq}\lesssim \tau_{\rm cool}$ translates into
\begin{equation} \label{eqRut}
	2\Gamma_{\rm Rut}\gtrsim u_0^{3}\,\Gamma_{\rm Comp}+u_0^{2}\,\Gamma_{\rm brem} \, ,
\end{equation}
and at high redshift, where Compton cooling is dominant, reads
\begin{equation} \label{eqcondition}
	1>\frac{\Gamma_{\rm Comp}}{2\Gamma_{\rm Rut}}u_0^{3}\sim\frac{\rho_\gamma}{\rho_{\rm DDM}^0}\left(\frac{T_{\rm vir}}{m_{p'}}\right)^{3/2}\left(\frac{m_{p'}}{m_{e'}}\right)^{7/2}\sim10^{-14}\left(\frac{m_{p'}}{m_{e'}}\right)^{7/2} \, .
\end{equation}
Here we have used the values of the virial temperature and DDM density given in Eqns.~\eqref{eq:T_vir} and~\eqref{eq:rho_0}. It can be shown that the requirement, Eq.~\eqref{eqcondition}, does not hold within the relevant parameter regions for a simple model with only $p'$, $e'$ and a dark photon. Specifically, we find that Eq.~\eqref{eqcondition} is satisfied only in the region of parameter space where $\tau_{\rm cool} > t_{\rm uni}$, in which case the substructure formation is too slow and the dark accretion mechanism is inefficient. Thus, a massless dark photon alone as a mediator is insufficient to provide an equilibration rate which is fast enough to comply simultaneously with all other requirements. However, if there is, for example, an additional interaction which is sufficiently short ranged so that it does not affect the accretion process, but strong enough to ensure fast equilibrium, then a much shorter $\tau_{\rm eq}$ can be achieved.
We limit the interaction range by considering an additional mediator, $\phi$, heavier then the (inverse) sub-structure size. In this case the revised rate becomes,
\begin{equation} \label{Gammaphi}
		\Gamma_{\rm Rut}\to\Gamma_{\phi}\simeq\frac{64\sqrt{2 \pi^5}\,n}{m_{p'}m_{e'}}\log\left(\frac{8\,m_{e'}^2u_{0}^2}{m_\phi^2}\right)\,,
\end{equation}
where in the above we assumed for simplicity $m_\phi \ll m_{e^\prime} u_0$\footnote{This assumption is not a necessary requirement. It is only taken here to provide an expression for the rate, Eq.~\eqref{Gammaphi}, analogous to Eq.~\eqref{GammaRut}. Once relaxed, the form of Eq.~\eqref{Gammaphi} changes, but the overall effect on the result is negligible.}.
Here, we have set the mediator's coupling to $\alpha_\phi=4 \pi$, the maximal value allowed by unitarity, and replaced the squared dark photon Debye mass ($\simeq 4 \pi \alpha^\prime n / T_{\rm vir}$) within the log, by $m_\phi^2$.  Thus, the condition of Eq.~\eqref{eqRut} becomes
\begin{equation}\label{eq:eq_condition}
	2\,\Gamma_{\phi}\gtrsim u_0^{3}\,\Gamma_{\rm Comp}+u_0^{2}\,\Gamma_{\rm brem}.
\end{equation}

\subsubsection{Large Duty Cycles}
\label{subsec:Duty_Cycles}

Accretion in the hidden sector occurs only when the inequality $\tau_{\rm eq}<\tau_{\rm cool}<t_{\rm univ}$ holds. 
The time it takes to accrete a substructure of mass $M_{\rm DDM}^{\rm sub}$ onto the BH can be obtained by integrating Eq.~\eqref{eq:MdotBH_vh}, neglecting the contribution from the visible sector.
Doing so, one obtains,
\begin{equation}
\tau_{acc}\approx\frac{\eta}{1- \eta}\frac{\tau_{Sal}}{\zeta}\frac{L_{\rm Edd}^h}{L^h}\log\left[1+\frac{\rho_{\rm DDM}^0}{\bar{\rho}}\left(\frac{M_{\rm BH}}{M_{\rm vir}}\right)^2\left(\frac{T_{\rm vir}}{T_f}\right)^3\right].
\end{equation}

If accretion is faster than cooling, the substructure will not have enough time to form before being accreted, in which case accretion cannot be continuous.  Therefore, the hidden duty cycle is approximately the ratio of $\tau_{\rm acc}$ and $\tau_{\rm cool}$ with a maximal value of unity. We define this duty cycle as
\begin{equation}\label{eq:hidden_duty}
D^h=\min\left[1,\frac{\tau_{\rm acc}}{\tau_{\rm cool}}\right] \Theta(2\,\Gamma_{\phi}- u_0^{3}\,\Gamma_{\rm Comp}-u_0^{2}\,\Gamma_{\rm brem})\Theta(t_{\rm uni} - \tau_{\rm cool}) \, ,
\end{equation}
where the step functions enforce the conditions $\tau_{\rm eq}<\tau_{\rm cool}<t_{\rm univ}$.
This expression depends on the BH mass through $\tau_{\rm acc}$, and on time through the $z$ dependence of $t_{\rm Comp}$. Because of this, when the expression for $D^h$ is plugged into Eq.~\eqref{eq:MdotBH_vh}, the equation can no longer be solved analytically as was possible for a constant $D^h$.  Solving the integral numerically, we obtain additional constraints on the parameter space, which are more stringent than those obtained earlier following Eq.~\eqref{eq:Hidden_Accretion}.

\subsection{Allowed Parameter Space}

\begin{figure}
\begin{center}
\includegraphics[width=0.47\textwidth]{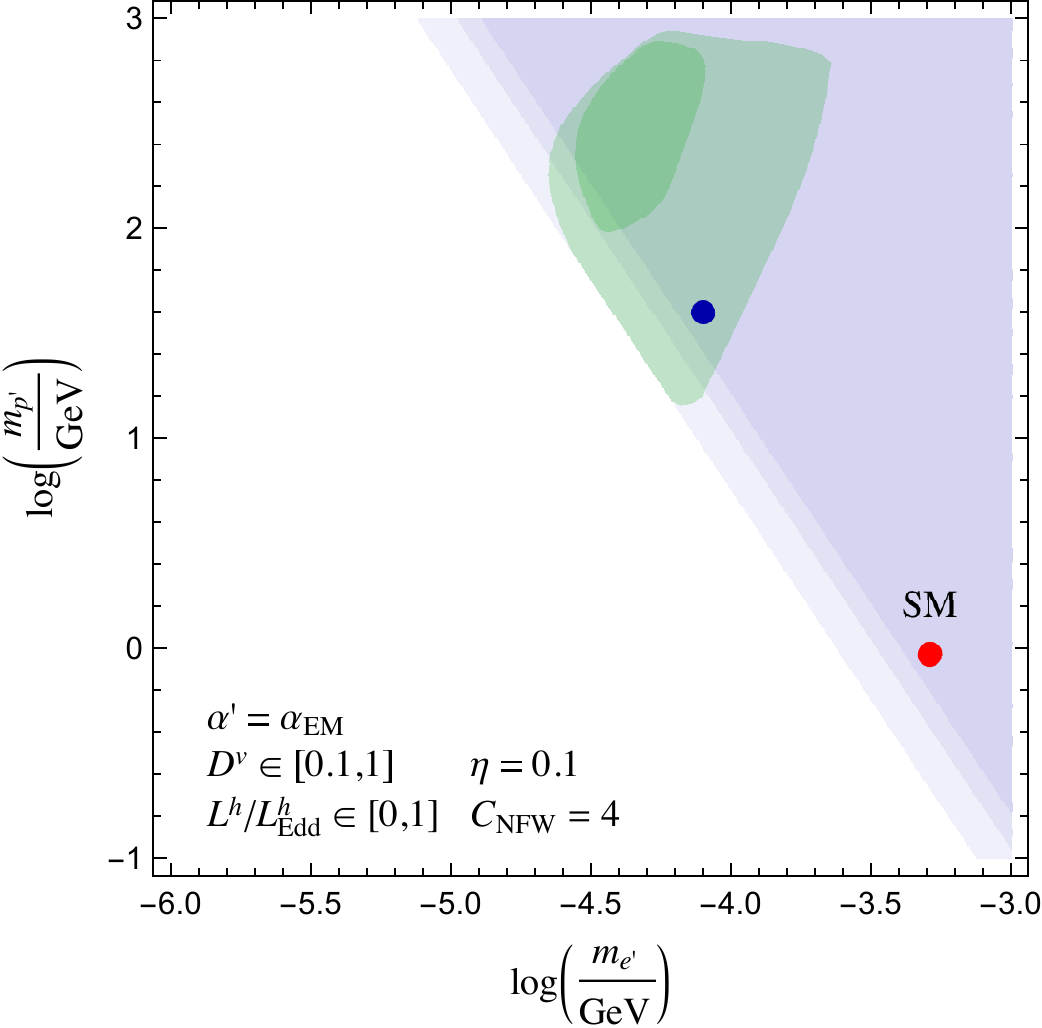}\;\;\;\;\includegraphics[width=0.47\textwidth]{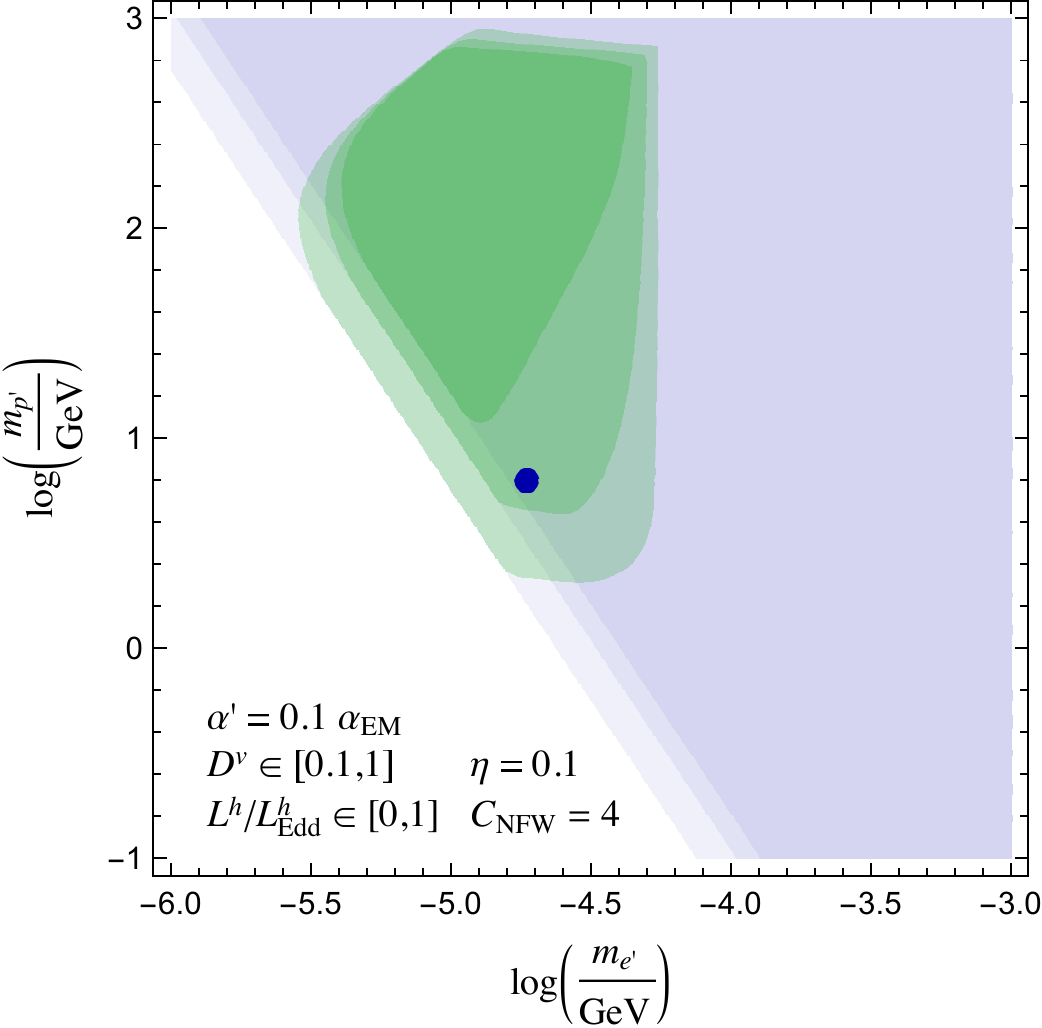}
\caption{\label{FIG:simple_region} 
Consistent region of parameters for a characteristic choice of $\eta=0.1$ and $C_{\rm NFW}=4$. \textbf{Left panel}: The dark coupling is set to $\alpha^\prime=\alpha_{\rm EM}$. The \textit{red dot} represents the SM values ($\alpha_{\rm EM}, m_e, m_p$). \textbf{Right panel}: The dark coupling is set to $\alpha^\prime=0.1 \alpha_{\rm EM}$. In the \textit{shaded blue regions} the tension in the measured SMBH mass presented in Sec.~\ref{sec:AGNs} is resolved under the naive assumptions in regards to the presence of an accretion disk discussed in Sec.~\ref{sec:naive}, while in the \textit{shaded green regions} the conditions to form an accretion disk presented in Sec.~\ref{sec:conditions} are also met. The \textit{increasingly opaque blue and green regions} represent the regions in which we assumed different maximal seed BH mass: lightest for  $M_\mathrm{seed}=10^6~ M_\odot$, medium for $M_\mathrm{seed}=10^4~ M_\odot$ and darkest for $M_\mathrm{seed}=10^2~ M_\odot$.
The \textit{blue dots} represent the values used to calculate the BH mass evolution shown on the left and right panels of Fig.~\ref{FIG:money} respectively.}
\end{center}
\end{figure}

\begin{figure}
\begin{center}
\includegraphics[width=.47\textwidth]{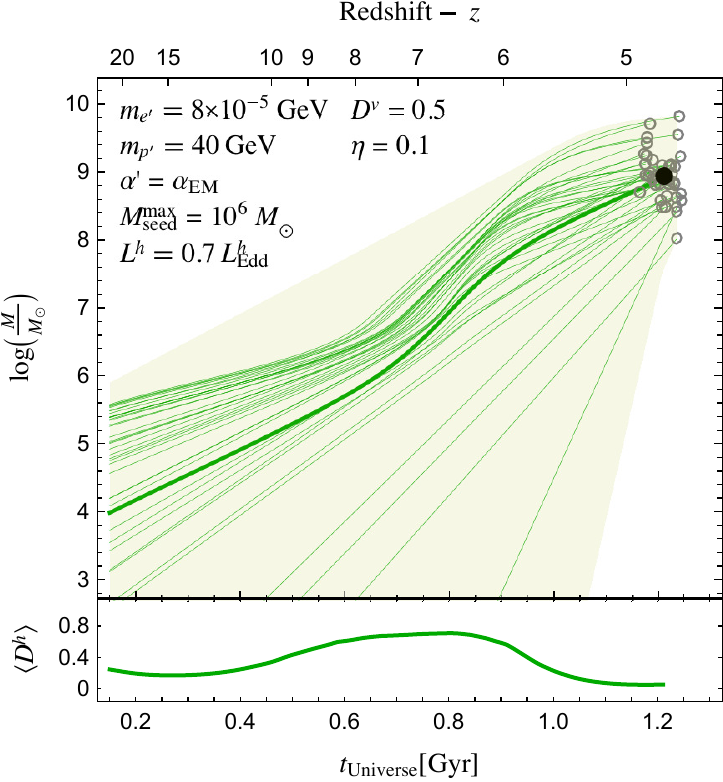}\;\;\;\;\includegraphics[width=.47\textwidth]{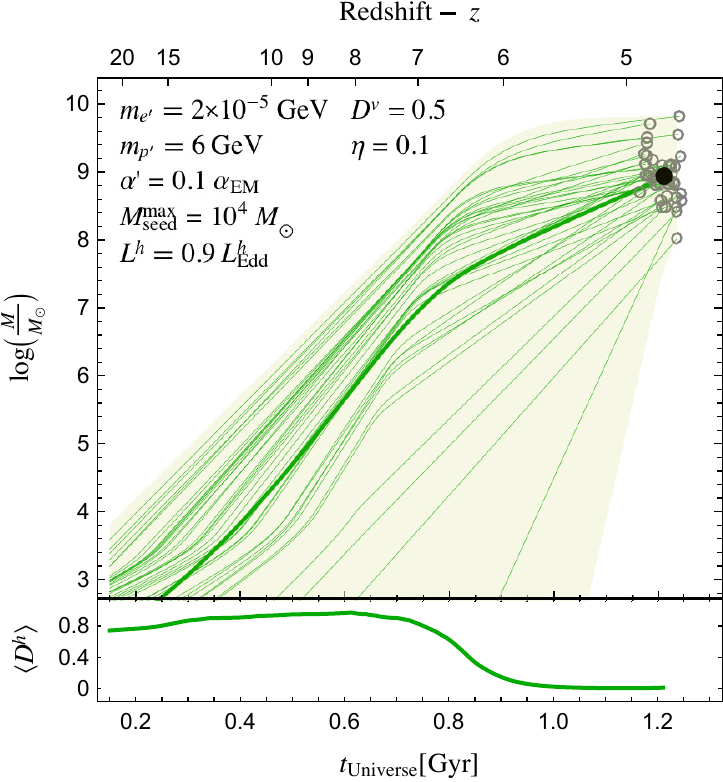}
\caption{\label{FIG:money} The SMBH mass as a function of cosmic time, including the effect of DDM accretion, demonstrated for two sets of parameter values from the preferred green shaded regions of Fig.~\ref{FIG:simple_region}. \textbf{Left panel}: Corresponds to the values indicated by the blue dot in the left panel of Fig.~\ref{FIG:simple_region}, for the choice of $L^h=0.7~ L_{\rm Edd}^h$ and for a maximal seed BH mass of $10^6 M_\odot$ . \textbf{Right panel}: Corresponds to the values indicated by the blue dot in the right panel of Fig.~\ref{FIG:simple_region}, for the choice of $L^h=0.9~ L_{\rm Edd}^h$ and for a maximal seed BH mass of $10^4 M_\odot$. The \textit{gray circles} represent the measured SMBH masses from the AGN sample presented and analyzed in \cite{Trakhtenbrot:2010hj}, the \textit{black dots} represent the sample mean and the \textit{thin green curves} show the regression of each SMBH mass back to $z\sim 20$. The curves have been calculated assuming $\eta=0.1$ and a visible duty cycle $D=0.5$. The \textit{thick green curves} show the regression in time of the sample's mean mass. The \textit{yellow shaded region} in both panels corresponds to the region that could be covered by allowing the visible duty cycle to vary between $0.1$ and $1$, while allowing for the same maximal seed BH masses as for the curves in each panel. \textbf{Bottom panels}: The sample's mean value of the hidden duty cycle, Eq.~\eqref{eq:hidden_duty}, as a function of cosmic time, for the same parameters as in the top panels. Hidden accretion is quenched whenever this value goes to zero.}
\end{center}
\end{figure}

We show the results of our analysis in Fig.~\ref{FIG:simple_region} for the three parameters of our model, $\{ m_{e^\prime}, m_{p^\prime}, \alpha^\prime \}$ where we have taken the mediator mass to be its minimal allowed value, $m_\phi=(R_{\rm DDM}^{\rm sub})^{-1}$, and the coupling to be maximal, $\alpha_\phi=4\pi$.  In the shaded blue regions, DDM accretion together with baryonic matter accretion can account for all the 40 BH 
masses of Ref.~\cite{Trakhtenbrot:2010hj} measured at redshift $z=4.8$, starting from various maximal seed BH masses between $10^2 - 10^6 \ M_\odot$ at $z = 20$ (different opacities correspond to different seed masses), following the analysis we described in Section~\ref{sec:naive}. However, this simple analysis  is based on the assumption that a dark accretion disk exists continuously, with enough DDM at its disposal to fuel the BH growth. Once we take into account the conditions to form and sustain a dark accretion disk, as discussed in Sec.~\ref{sec:conditions}, the available parameter space shrinks from the blue to the green shaded regions (again with the varying opacities corresponding to different maximal seed masses). The green region is bounded at large $m_{e'}$ as a result of the requirement that the cooling time scale should be shorter than the age of the universe, and at small $m_{e'}$  by the lower bound of the additional mediator mass ($m_\phi\gtrsim (R_{\rm DDM}^{\rm sub})^{-1}$).

It is interesting to note that the SM values of $m_{e},~m_{p}$ and $\alpha$, marked by the red dot on the left panel of Fig.~\ref{FIG:simple_region}, are within the region consistent with the naive analysis following Eq.~(\ref{eq:Hidden_Accretion}), but outside of the region which takes into account the additional requirements to form an accretion disk. In other words, using our criteria, the SM would not efficiently form an accretion disk.  This could indicate that other, more complicated, processes (such as turbulences, magnetohydrodynamics, etc.) could be important in   our analysis. Thus, in a more realistic scenario, blue-but-not-green regions in Fig.~\ref{FIG:simple_region} might well allow for efficient accretion.

Fig.~\ref{FIG:money} shows the time evolution of the SMBH mass including the effect of DDM accretion for the sample of SMBHs  considered in this study. This plot should be compared with the left panel of Fig.~\ref{FIG:Netzer} where the effects of DDM accretion are not included. Each panel of Fig.~\ref{FIG:money} corresponds to different values of the parameters $\{ m_{e^\prime}, m_{p^\prime}, \alpha^\prime \}$ (again taking the minimal $m_\phi$ and maximal $\alpha_\phi$),  a different maximal $M_{\rm seed}$ and the curves correspond to a chosen value of $L^h/L^h_{\rm Edd}$. In each panel, the yellow shaded region corresponds to the region where $L^h/L^h_{\rm Edd}$ can vary between 0 and 1, the visible duty cycle can vary between 0.1 and 1 and the maximal seed mass is held fixed.  This should be understood as the region that corresponds to all possible evolutions of SMBHs in the sample, from some maximal seed, under the assumptions described. Clearly, the addition of DDM to the accretion history allows for many evolution paths that reach the measured final mass without requiring too large seeds or super-Eddington accretion. The bottom panels present the (sample averaged) dark duty cycle as a function of cosmic time. Evidently, for some DDM parameters, DDM accretion occurs early in cosmological time, essentially turning off later on (right panel), while for other parameters, DDM accretion ignites at late times and contributes to later fast growth (left panel). Observations, such as those described in Sec.~\ref{sec:AGNs}, point towards a scenario more similar to the former case, i.e. early fast growth. We learn that DDM accretion, as described in this study, could provide an explanation for this early growth phase.

\section{Conclusions} 
\label{sec:Conclusions}
We have performed a preliminary study aimed at understanding whether DDM could form an accretion disk, and alleviate tension regarding observations of SMBHs with masses as large as $10^9 \ M_\odot$ at early redshifts.  Our results are based on the critical assumption that a hidden sector which is similar enough to the visible sector can allow for a viscous disk which accretes onto a SMBH once formed. Under this assumption, we find that there is a region of the parameter space that allows DDM to cool and collapse down to the scale of the accretion disk and could yield a continuous and exponential growth of the SMBH, with fast cooling and efficient accretion. We find that for some regions in the DDM parameter space, the accretion history for the sample of SMBHs considered, is more consistent with the 
observation of early fast accretion that slows down at later times. This result is evident in the right panel of Fig.~\ref{FIG:money} which should be compared with the left panel of Fig.~\ref{FIG:Netzer}, where the effects of DDM accretion are not included. Our results complement the parameter space of Ref.~\cite{Fan:2013yva}, where the authors studied the formation of a galactic DDM disk (at galactic scales) with the same simplified DDM model.

A future dedicated numerical simulation, beyond the scope of this paper, could provide more insight into our results. However, such a numerical study would require sub-parsec resolution, beyond the capabilities of current simulations. Furthermore, an additional future study could include observationally searching for signatures of DDM accretion in the form of an anomalous AGN-like spectrum from a seemingly dormant galaxy. This could occur if kinetic mixing between the dark photon and the visible photon exists, and for a SMBH which is observed while accreting DDM but no baryonic matter (or where some of the visible light is blocked because of the galaxy's orientation). We leave this avenue for a future study.

{\it\bf Acknowledgments:} 
We thank Avishai Dekel, Amir Levinson, Hagai Netzer, Lisa Randall and Jakub Scholtz for enlightening discussions.  NJO is grateful to the Azrieli Foundation for the award of an Azrieli Fellowship. NJO and TV are supported in part by the I-CORE Program of the Planning Budgeting Committee and the Israel Science Foundation (grant No. 1937/12). TV is further supported by the Israel Science Foundation-NSFC (grant No. 2522/17) and by the German-Israeli Foundation (grant No. I-1283- 303.7/2014), by the Binational Science Foundation (grant No. 2016153) and by a grant from the Ambrose Monell Foundation, given by the Institute for Advanced Study. OS acknowledges support from the Clore Foundation. LU acknowledges support from the PRIN project ``Search for the Fundamental Laws and Constituents'' (2015P5SBHT\textunderscore 002).

\appendix

\section{A derivation of the initial DDM profile} 
\label{sec:ddm_profile_derivation}

In section~\ref{sub:galactic_toy_model} it was assumed that at early redshifts of $z\sim 20 - 30$, i.e. before the formation of an accretion disk, the DDM component was spherically symmetric and in hydrostatic equilibrium within the gravitational potential of the CDM component. Such a system is described by the following two equations,
\begin{align}
\vec{\nabla}P & = -\rho_{\rm DDM}(\vec{r}) \vec{\nabla} \Phi \, ,\label{eq:hydroeq} \\
\nabla^2 \Phi & = 4\pi G \left( \rho_{\rm CDM}(\vec{r}) + \rho_{\rm DDM}(\vec{r}) \right) \, .
\label{eq:Poisson}
\end{align}
Here $\Phi (r)$ is the Newtonian gravitational potential, $P = \rho_{\rm DDM} T/\mu m_{p^\prime}$ is the pressure of the DDM gas with $T$ its temperature, and $\mu = \rho_{\rm DDM}/{m_{p^\prime} n_{\rm DDM}} \approx \rho_{p^\prime}/{(n_{p^\prime}+n_{e^\prime}) m_{p^\prime}} = 1/2$.

\begin{figure}[t]
\begin{center}
\includegraphics[width=.48\textwidth]{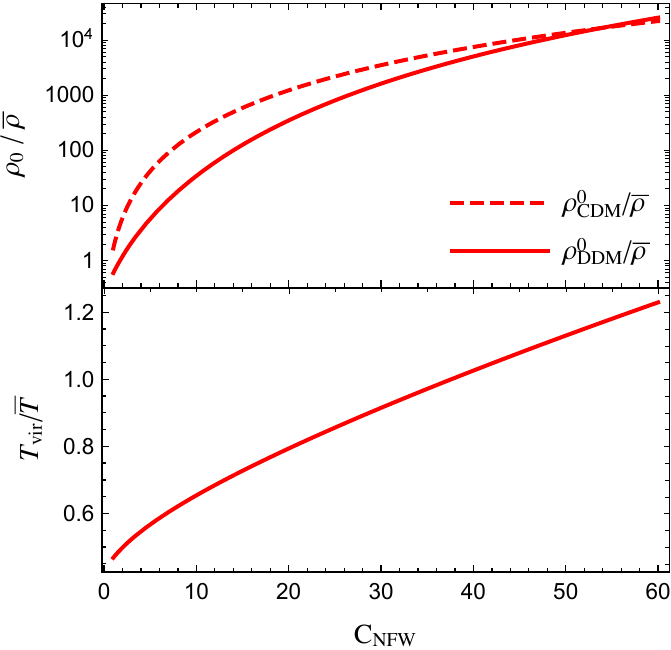}\;\;\;\;\;\includegraphics[width=.48\textwidth]{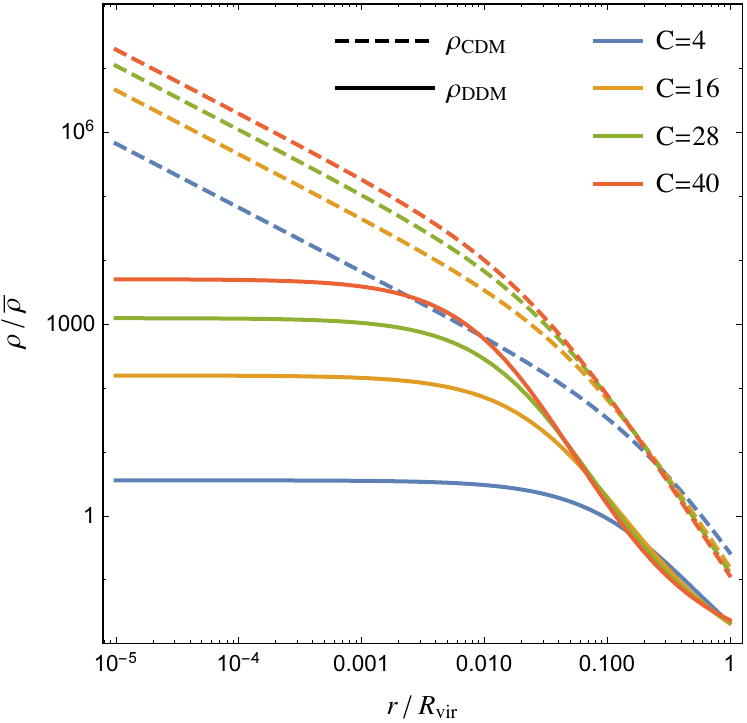}
\caption{\label{FIG:rho_vs_C} 
{\bf Left panel}: The normalization of the CDM and DDM density profiles $\rho_{\rm CDM}^0/\bar{\rho}$ and $\rho_{\rm DDM}^0/\bar{\rho}$ ({\bf top}), and the virial temperature in units of $\bar T$ as defined below Eq.~\eqref{eq:b_def} ({\bf bottom}), each as functions of the concentration parameter, $C\equiv R_{\rm vir}/R_s$,  as defined below Eq.~\eqref{eq:NFWnorm}. Note that while $\rho_{\rm DDM}^0$ is the value of the DDM profile at the origin, $\rho_{\rm CDM}^0$ is the normalization of the \textit{divergent} NFW profile, corresponding to its value at around $r\approx 0.47~ R_s$.
{\bf Right panel}: The DDM ({\it solid curves}) and CDM ({\it dashed curves}) density profiles as functions of the radius, for different values of the concentration parameter $C$.}

\end{center}
\end{figure}

In this appendix we provide an alternative derivation of the DDM gas profile, compared to that typically found in textbooks, see e.g. Ref.~\cite{Mo:2010gfe}. Unlike other derivations in the literature, the derivation below also offers a self consistent calculation of the virial temperature of the isothermal sphere.
We consider an isothermal profile for the DDM, parametrized as
\beq
\rho_{\rm DDM}(r) = \tilde\rho \exp\left(-\beta \Phi(r)\right) \, ,
\eeq
where $\beta = \frac{\mu m_{p^\prime}}{T}$. 
Note that the freedom to transform $\Phi\to \Phi+$constant is retained so long as one changes the value of $\tilde\rho$ accordingly.
We impose the normalization condition,
\begin{equation}\label{eq:DDM_mass}
	M_{\rm DDM}^{\rm gal}=\tilde{\rho}\int d^3r e^{-\beta \Phi(r)}.
\end{equation}
Choosing the gauge $\Phi(\infty)=0$, one can calculate the gravitational energy stored within a DDM sphere,
\begin{equation}\label{eq:e_grav} 
-E_{\rm grav}=\int d^3r \,\rho_{\rm DDM}\Phi=\int d^3r \,\tilde{\rho}e^{-\beta \Phi}\Phi=-\tilde{\rho}\frac{\partial}{\partial \beta}\int d^3r e^{-\beta \Phi}=M_{\rm DDM}^{\rm gal}\frac{1}{\tilde{\rho}}\frac{\partial \tilde{\rho}}{\partial \beta} \, ,
\end{equation}
where in the last equality we have used Eq.~\eqref{eq:DDM_mass}. Assuming the system to be virialized, we have,
\begin{equation} \label{eq:virial}
	\frac{3M_{\rm DDM}^{\rm gal}}{2 \beta}=\frac{3}{2}N_{\rm DDM}T_{\rm vir}=\frac{1}{2}E_{\rm grav},
\end{equation}
where $N_{\rm DDM}=M_{\rm DDM}^{\rm gal}/\mu m_{p^\prime}$ is the number of DDM particles within the galaxy. From Eqns.~\eqref{eq:e_grav} and \eqref{eq:virial} we conclude that the normalization of the isothermal sphere satisfies,
\begin{equation}\label{eq:dln_rho_dln_beta}
 	\frac{\rm d \log \tilde{\rho}}{\rm d\log \beta}=-3.
 \end{equation} 
Together with appropriate boundary conditions, Eqns.~\eqref{eq:Poisson},~\eqref{eq:DDM_mass} and \eqref{eq:dln_rho_dln_beta} uniquely determine $\Phi$, $\beta$ and $\rho_{\rm DDM}^0$. We specify the following boundary conditions,
\begin{equation}
	\Phi(\infty)=0\;\;,\;\;\Phi'(r)=\frac{G M(r)}{r^2} \, ,
\end{equation}
the latter dictated by Newton's second law.
In the physical system of interest, it may be assumed that $\rho_{\rm DDM}\ll \rho_{\rm CDM}$. Self consistency is demonstrated in the right plane of Fig.~\ref{FIG:rho_vs_C}.
We further assume that both $\rho_{\rm DDM}$ and $\rho_{\rm CDM}$ vanish for $r>R_{\rm vir}$. Solving Eq.~\eqref{eq:Poisson} under these assumptions, with an NFW profile, Eq.~\eqref{eq:NFW}, for the CDM, one finds,
\begin{equation}\label{eq:ddm_profile}
	\rho_{\rm DDM}(r) =\tilde{\rho}\exp\left(-\frac{b}{1+C}\right)\left(1+\frac{r}{R_s}\right)^{b R_s/r}.
\end{equation}
Here $C$ is the NFW concentration parameter defined below Eq.~\eqref{eq:NFWnorm}, and $b$, defined in Eq.~\eqref{eq:b_def}, can be rewritten as
\begin{equation}
b = \frac{3}{C^{2}}\frac{\rho_{\rm CDM}^0}{\bar \rho} \frac{\bar T}{T_{\rm vir}} \, , \label{eq:bapp}
\end{equation}
where $\bar \rho = 3 M^{\rm gal}_{\rm DM}/{4\pi R_{\rm vir}^3}$, the virial temperature is defined by Eq. \eqref{eq:virial} and we define a temperature scale $\bar{T}\equiv{G M^{\rm gal}_{\rm DM} m_{p^\prime}}/{2 R_{\rm vir}}$. Note that taking $r\to 0$, the DDM central density is given by
\begin{equation}
	\rho_{\rm DDM}^0\equiv \lim_{r\to 0}\rho_{\rm DDM}(r)=\exp\left(\frac{C~b}{1+C}\right)\tilde{\rho} \, .
\end{equation}
Eqns.~\eqref{eq:DDM_mass} and~\eqref{eq:dln_rho_dln_beta} provide two relations that allow one to determine $T_{\rm vir}$ and $\rho_{\rm DDM}^0$, leaving $C$ as the only free parameter of the DDM profile. The derived temperature and profile normalizations are shown in Fig.~\ref{FIG:rho_vs_C}.

\section{Temperature Evolution} 
\label{sec:temperature_evolution}
For an accretion disk to form, it is necessary to cool the hidden electrons and hidden protons sufficiently fast.   In this appendix we study the conditions for this to happen. 

Assuming the number density of electrons and protons is constant in time, the equation governing the temperature evolution can be obtained from the Boltzmann equations for the energy densities. The dissipative hidden sector introduced in Sec.~\ref{sec:Model} allows for cooling through bremsstrahlung and through Compton scattering with the hidden CMB.  Moreover, the mass hierarchy $m_{e'} \ll m_{p'}$ ensures that only electrons are efficiently directly cooled. The protons can then follow the electron's thermal bath via Rutherford scattering. Neglecting the expansion of the universe the Boltzmann equations read
\begin{align}
\label{eq:BEs}
	&\dot T_{e'}=\frac{2}{3}\dot Q_{\rm Rut}-\Gamma_{\rm Comp}(T_{e}-T_{\gamma})-\Gamma_{\rm brem}m_{e'}v_{\rm th}\,, \\
	&\dot T_{p'}=-\frac{2}{3}\dot Q_{\rm Rut}\,,
\end{align}
where $v_{\rm th}^2=T_{e'}/m_{e'}+T_{p'}/m_{p'}$. The Compton and bremsstrahlung rates are given simply by,
\begin{equation}
	\Gamma_{\rm Comp}=\frac{8 \pi}{45}\frac{\sigma_{T}'}{m_{e'}}T_{\gamma}^4\;\;,\;\;\Gamma_{\rm brem}=\alpha'n\sigma_{T}'\,,
\end{equation}
The heat transfer rate due to Rutherford scattering is given by (see for example the appendices of Refs.~\cite{Munoz:2015bca,Barkana:2018qrx})
\begin{equation}\label{eq:qdot}
	\frac{2}{3}\dot Q_{\rm Rut}=\frac{m_{e'}n \sigma_{T}'}{\sqrt{8 \pi}m_{p'}v_{\rm th}^3}\log\left(\frac{32m_{e'}^2T_{p'}^2v_{\rm th}^4}{3 \pi n^2 \sigma_{T}'e^{2+2\gamma}}\right)(T_{p'}-T_{\rm e'})\equiv \Gamma_{\rm Rut}(T_{p'}-T_{e'})v_{\rm th}^{-3},
\end{equation}
where $\gamma$ is the Euler–Mascheroni constant and we have assumed that the IR divergence of the Rutherford scattering is regularized by the Debye mass $\simeq \sqrt{4\pi\alpha'n/T_{p'}}$.

For the purpose of this study, we assume that equilibrium is reached much faster than the cooling time of the DDM gas, in which case the cooling timescale is given by,
\begin{equation}
	\tau_{\rm cool}=\left|\frac{T_{p'}+T_{e'}}{\dot T_{p'}+\dot T_{e'}}\right|_{T_{p'}=T_{e'}=T_{\rm vir}}=\frac{2}{\Gamma_{\rm Comp}+u_0^{-1}\Gamma_{\rm brem}}.
\end{equation}
The equilibration time can be estimated as the time it takes for the electron bath to reach $T_{e'}^*$, the temperature at which the equilibration rate becomes sizable, i.e.~when the first term on the RHS of Eq.~\eqref{eq:BEs} is of order the sum of the other two terms.

While this timescale is easily calculated numerically, it is constructive to derive an analytical expression.  As we dial up the equilibration rate, the corresponding timescale, $\tau_{\rm eq}$ drops and therefore  $T_{e}^*\rightarrow T_{\rm vir}$.  In this limit one finds,
\begin{equation}
\tau_{\rm eq} \equiv \left| \frac{T_{\rm vir} - T_{e}^*}{\dot T_{e}}\right|_{T_{p'}=T_{e'}=T_{\rm vir}} \simeq \frac{\Gamma_{\rm Comp}u_0^3+\Gamma_{\rm brem}u_0^2}{2\Gamma_{\rm Rut}}\, \tau_{\rm cool}\,,
\end{equation}
and therefore for the protons  to cool sufficiently fast, one needs $\Gamma_{\rm Comp} u_0^3+\Gamma_{\rm brem} u_0^2 \lesssim 2\Gamma_{\rm Rut}$. 
Since the LHS of this expression is proportional to the ratio of dark photon to DDM energy density, it is large in much of the parameter space. Indeed, by using the values for the virial temperature and DDM density obtained in App.~\ref{sec:ddm_profile_derivation}, it is straightforward to check that the conditions of Sec.~\ref{sec:DMAccretion} are not satisfied if equilibration is dominated by scattering via the same hidden photons that are responsible for the radiation of the accretion disk. In this case, an additional (possibly short-range) coupling between the hidden electrons and protons is necessary.

\bibliographystyle{JHEP}
\bibliography{AGNs.bib}

\end{document}